\documentclass[twocolumn,superscriptaddress, graphics,floatfix,nofootinbib,
tightenlines,nobibnotes,aps,prb]{revtex4-1}

\usepackage{color}
\usepackage{bm}
\usepackage{amsmath,units}
\usepackage{amssymb}
\usepackage{graphicx,hyperref}

\makeatletter
\@ifundefined{textcolor}{}
{%
 \definecolor{BLACK}{gray}{0}
 \definecolor{WHITE}{gray}{1}
 \definecolor{RED}{rgb}{1,0,0}
 \definecolor{GREEN}{rgb}{0,1,0}
 \definecolor{BLUE}{rgb}{0,0,1}
 \definecolor{CYAN}{cmyk}{1,0,0,0}
 \definecolor{MAGENTA}{cmyk}{0,1,0,0}
 \definecolor{YELLOW}{cmyk}{0,0,1,0}
}

\usepackage{amsfonts}%\usepackage{paralist}              	%f�r Formatierung der Aufz�hlung
\usepackage{array}\usepackage{bm}\usepackage{dsfont}\usepackage{ulem}

\newcommand{\bs}[1]{\boldsymbol{#1}}

\renewcommand{\vec}[1]{\textbf{\textit{#1}}}
\newcommand{\mx}[1]{\ensuremath{\left( \begin{matrix} #1 \end{matrix} \right) }}
\newcommand{\zref}[1]{(\ref{#1})}

\newcommand{\s}{^{\text{s}}}
\newcommand{\abs}[1]{|#1|}
\newcommand{\verweis}[1]{\ensuremath{\stackrel{(\ref{#1})}{=}}}
\def\R{\vec{R}}

\makeatother

\begin{document}

\title{Elastoconductivity as a probe of broken mirror symmetries}

\author{Patrik Hlobil}

\affiliation{Karlsruher Institut f\"{u}r Technologie, Institut f\"{u}r Theorie der Kondensierten
Materie, 76128 Karlsruhe, Germany}

\author{Akash V. Maharaj}

\affiliation{Department of Physics, Stanford University, Stanford, California
94305, USA}
\affiliation{Stanford Institute for Materials and Energy Sciences, SLAC National Accelerator Laboratory,
2575 Sand Hill Road, Menlo Park, CA 94025, USA }

\author{Pavan Hosur}

\affiliation{Department of Physics, Stanford University, Stanford, California
94305, USA}

\author{M.C. Shapiro}

\affiliation{Stanford Institute for Materials and Energy Sciences, SLAC National Accelerator Laboratory,
2575 Sand Hill Road, Menlo Park, CA 94025, USA }

\affiliation{Department of Applied Physics, Stanford University, Stanford, California
94305, USA}

\author{I.R. Fisher}

\affiliation{Stanford Institute for Materials and Energy Sciences, SLAC National Accelerator Laboratory,
2575 Sand Hill Road, Menlo Park, CA 94025, USA }

\affiliation{Department of Applied Physics, Stanford University, Stanford, California
94305, USA}

\author{S. Raghu}

\affiliation{Department of Physics, Stanford University, Stanford, California
94305, USA}

\affiliation{Stanford Institute for Materials and Energy Sciences, SLAC National Accelerator Laboratory,
2575 Sand Hill Road, Menlo Park, CA 94025, USA }

\date{\today }

\begin{abstract}
We propose the possible detection of broken mirror symmetries in correlated
two-dimensional materials by elastotransport measurements. Using linear
response theory we calculate the``shear conductivity" $\Gamma_{xx,xy}$,
defined as the linear change of the longitudinal conductivity $\sigma_{xx}$
due to a shear strain $\epsilon_{xy}$. This quantity can only be
non-vanishing when in-plane mirror symmetries are broken and we discuss
how candidate states in the cuprate pseudogap regime (e.g. various
loop current or charge orders) may exhibit a finite shear conductivity.
We also provide a realistic experimental protocol for detecting such
a response.
\end{abstract}

\maketitle

\section{Introduction}

Phases of matter in solids are often empirically distinguished by their transport
properties. Electrical transport measurements probe long wavelength
properties of the system, and in metallic phases, they are sensitive
to electronic excitations near the Fermi level. In addition, these measurements 
often exhibit singular features at phase transitions, thus indicating the onset of broken symmetries. 
However, aside from a few instances such as the anomalous Hall effect in ferromagnets,
the exact form of the broken symmetry is not usually evident from a transport 
measurement. Experiments that aid in directly identifying subtle forms of broken symmetry are 
therefore invaluable in the study of strongly correlated electron materials.

Motivated by such considerations, we study the \textit{linear} change
of electrical transport coefficients in the presence of applied strain
which we refer to as a ``shear conductivity''. If such
a linear response is present, it is necessarily encoded in a fourth-rank
tensor. As we discuss below, the shear conductivity is a binary indicator
of point group and mirror symmetry breaking and retains its character
as a transport coefficient in probing the dynamics of the quasiparticles
(or lack thereof) near the Fermi level. By contrast, ordinary transport
coefficients, which are second rank tensors, are at best indirect markers of 
such transitions.

Specifically, we discuss how a linear change in longitudinal conductivity
$\sigma_{xx}$ due to applied strain $\epsilon_{xy}$ can only occur
if certain vertical mirror plane symmetries are broken. When such symmetries
are absent a response of this sort is no longer forbidden, and is
therefore generically finite. 
While our considerations here are based solely on symmetry and are therefore quite general, we are primarily motivated by the cuprate superconductors, where a variety of broken symmetry
phases are likely present in the pseudogap regime of hole-doped materials.
Several candidate order parameter theories have been proposed for
the pseudogap regime including current-loop phases\cite{Varma99Pseudogap,Varma2006},
$d$-density wave phases\cite{chakravarty2001hidden,laughlin2014fermi}, various forms of charge
order\cite{zaanen1989charged,machida1989magnetism,emery1993frustrated,emery1999stripe,white1998density,sachdev2013bond,efetov2013pseudogap, wang2014charge,chowdhury2014Density,Maharaj2014}, electron nematic phases\cite{nie2014quenched,wang2014charge}, and pair density wave states\cite{berg2009theory, lee2014amperean,Agterberg2015emergent}
to name just a few. 

Here, we have focussed on two phases that break among other symmetries, point group and mirror symmetry.  These are the variants of phases with loop current order as well as those with charge order.   Our key result is that there is a finite and measurable shear conductivity in both states.  Furthermore,  we predict a parametrically higher shear conductivity response in the orbital current loop phase near its onset temperature, when compared to the response of charge ordered states.
Our analysis was inspired by an elegant set of experiments~\cite{Chu12,Kuo1,Kuo2,Riggs}
that have utilized transport measurements in the presence of strain as a probe 
of nematicity.
Our goal is to generalize these experimental protocols
to help uncover more subtle patterns of symmetry breaking. We also note that while our
focus here is on electrical transport coefficients, the symmetry
considerations discussed below apply equally to other measurements,
such as ultrasound attenuation\cite{lupien2001ultrasound}, which also involves the determination
of a fourth rank tensor.

This paper is organized as follows: we first review the simple symmetry considerations which 
lead to a non-vanishing shear conductivity in Sec.~\ref{sec:symmetry}. We then discuss how this quantity is actually evaluated 
in the framework of the Kubo formula, before performing explicit calculations for model charge ordered systems and loop current phases in Sec.~\ref{sec:calculation}. Next, in Sec.~\ref{sec:training} we discuss how the response is trained in macroscopic crystals, before closing by discussing realistic experimental protocols for the measurement of the shear conductivity in Sec.~\ref{sec:measurements}.

\section{Symmetry considerations}\label{sec:symmetry}
 We define the  shear conductivity 
 as the elastoconductivity tensor component
$\Gamma_{xx,xy}\equiv\partial\sigma_{xx}/\partial\epsilon_{xy}$,
which describes the change of the longitudinal DC conductivity induced by
a shear strain of the crystal. The tensor $\Gamma_{xx,xy}$ can be non-vanishing \textit{only} when each
of the symmetries is broken: (i) reflection about the $xz$-plane, which has a normal vector along $y$: $\hat{\sigma}_{y}$
(ii) reflection about the $yz$-plane, $\hat{\sigma}_{x}$ and (iii) the combination
$\hat{\sigma}_{(x=y)}*C_{4}$. Here, $\hat{\sigma}_{(x=y)}$ denotes reflection about the vertical $(x=-y)z$
plane and $C_{4}$, a fourfold rotation about the principal $z$-axis. This follows from the fact that under each 
of these symmetry operations, $\sigma_{xx}$ is even while $\epsilon_{xy}$ is odd. 
More generally, in the presence of any of the 3 symmetries mentioned above, $\Gamma_{ij,kl}$ vanishes if the indices contain an odd number of
$x$ or $y$.  An example is $\Gamma_{xy,xx}$, which represents the change
in the Hall conductivity due to a longitudinal strain. On the other hand, inversion symmetry
imposes no restrictions $\Gamma_{ij,kl}$. Moreover, Onsager's
reciprocity theorem dictates that $\Gamma_{ij,kl}(M)=\Gamma_{ji,kl}(-M)$, where $M$ is odd under time-reversal (the final pair of indices is unaffected since strain is a symmetric time-reversal invariant tensor).  
Lastly, we remark that reciprocity does not relate  $\Gamma_{ij,kl}$ and
$\Gamma_{il,kj}$. Further details on symmetry properties of $\Gamma_{xx,xy}$ and the full elastotransport
tensor in a tetragonal system are provided in Appendix~\ref{secsym}.

\section{Elastoconductivity in model systems}\label{sec:calculation}
\subsection{Calculation of elastoconductivity}
Having disposed of generalities, we now compute the elastoconductivity tensor from an 
explicit microscopic model, using the Kubo formula, and taking into account the effects of strain.  
We consider a non-interacting Hamiltonian,  which captures the appropriate broken
symmetry, and takes the form 
\begin{align}
\mathcal{H}=\sum_{\vec{k},\alpha\beta}H_{\vec{k},\alpha\beta}c_{k,\alpha}^{\dagger}c_{k,\beta}=\sum_{\vec{k}}\hat{\Psi}_{\vec{k}}^{\dagger}\hat{H}_{\vec{k}}\hat{\Psi}_{k}\label{coe1},
\end{align}
where $\alpha,\beta$ can be spin, band or other quantum numbers,
$\Psi_{\vec{k}}^{\dagger}=(c_{\vec{k},1}^{\dagger},c_{\vec{k},2}^{\dagger},\ldots)$
denote the fermionic creation operators, and $\vec{k}$ is the crystal momentum. Anticipating the physical systems that we will apply
our results to in the next section, we restrict ourselves to $2D$;
however, the formalism generalizes in a straightforward way to 3D.
Without strain, the Kubo formula gives %
\footnote{Also the Hall conductivity can be calculated using the presented formalism.
Note that in this case one has to use a different response than \zref{coe3}. %
} 
\begin{align}
\sigma_{xx} & =-\pi e^{2}\int\limits _{\text{B.Z.}}\frac{d^{2}\vec{k}}{(2\pi)^{2}}\int\limits _{-\infty}^{\infty}dE\frac{\partial f(E)}{\partial E}\text{tr}\biggl[\bigl(\hat{A}_{\vec{k}}(E)\frac{\partial\hat{H}_{\vec{k}}}{\partial k_{x}}\bigr)^{2}\biggr]\label{coe2}
\end{align}
where $\hat{A}_{\vec{k}}(E)=\frac{i}{2\pi}\bigl[\hat{G}_{\vec{k}}(E+i/2\tau_{\text{sc}})-\hat{G}_{\vec{k}}(E-i/2\tau_{\text{sc}})\bigr]$
is the spectral function %
\footnote{Since we are calculating the DC conductivity, we have to assume a
finite scattering time $1/\tau_{\text{sc}}\ll\mu$, where $\mu$ is
the chemical potential, so that the spectral functions are sharply
peaked Lorentzians rather than delta functions.%
} and $f(E)$ is the Fermi distribution function. As explained in the
Appendix~\ref{sec:linearresponse}, the application of a strain $\hat{\epsilon}$
leads to a change of the Bravais lattice vectors $\{\vec{a}_{i}\}\rightarrow\{\bigl(\mathds1+\hat{\epsilon}\bigr)\vec{a}_{i}\}$,
which can easily be implemented in a tight-binding approach. This
has two main effects on a tight-binding Hamiltonian\cite{Pereira}:
(i) the tight-binding hopping parameters may change since they depend
on the distance between the atoms in general, and (ii) the momenta are modified 
according to $\vec{k}\rightarrow(\mathds1+\hat{\epsilon})\vec{k}$. As a consequence, 
the Hamiltonian is modified as $\hat{H}_{\vec{k}}\rightarrow\hat{H}_{(\mathds1+\hat{\epsilon})\vec{k}}^{s}$ where the superscript $s$ indicates the modified tight binding parameters, and the Brillouin zone is also altered accordingly. 

After introducing strain in \zref{coe2}, a coordinate transformation
$\vec{p}=(\mathds1+\hat{\epsilon})\vec{k}$ effectively undoes 
(ii) while introducing a Jacobian into the expression
for $\sigma_{xx}$. As a result, the DC conductivity in the strained
crystal can be written as 
\begin{align}
\sigma_{xx}^{\text s}= & -\frac{\pi e^{2}}{\text{det}(\mathds1+\hat{\epsilon})}\int\limits _{\text{B.Z.}}\frac{d^{2}\vec{p}}{(2\pi)^{2}}\int\limits _{-\infty}^{\infty}dE\frac{\partial f(E)}{\partial E}\nonumber \\
 & \text{tr}\biggl[\biggl(\hat{A}_{\vec{p}}^{s}(E)\bigl[(1+\epsilon_{xx})\frac{\partial\hat{H}_{\vec{p}}^{s}}{\partial p_{x}}+\epsilon_{xy}\frac{\partial\hat{H}_{\vec{p}}^{s}}{\partial p_{y}}\bigr]\biggr)^{2}\biggr]\label{coe3}
\end{align}
where the momentum integration spans the unstrained 1st Brillouin
zone of the lattice. The shear conductivity $\Gamma_{xx,xy}$, can
be computed from (\ref{coe3}) by setting $\epsilon_{xx}=\epsilon_{yy}=0$, expanding and extracting the linear coefficient via $\sigma_{xx}^{\text s}= \sigma_{xx} + \Gamma_{xx,xy} \epsilon_{xy} + \mathcal{O}(\epsilon_{xy}^2)$.

We now apply the formalism outlined above to study two candidate phases relevant to the pseudogap regime of the hole doped cuprates: (i) a particular form of charge order that breaks mirror symmetries, and (ii) two-dimensional loop current phases proposed first by Varma.

\subsection{Shear conductivity as a probe of charge order} \label{sec:chargeorder}
Despite the remarkable recent experimental progress in identifying (generally short range correlated) charge order
as a ubiquitous member of the cuprate phase diagram,\cite{tranquada1995evidence,hoffman2002four,howald2003Periodic,ghiringhelli2012long,chang2012direct,doiron2013hall} many fundamental
questions of symmetry remain. Candidate states such as predominantly
$d$-wave charge order\cite{sachdev2013bond,allais2014density,thomson2015charge} and criss-crossed stripes\cite{Maharaj2014} do
break the mirror symmetries required to produce a finite shear conductivity,
while conventional $s$-wave checkerboard charge density waves, or
unidirectional stripes\cite{zaanen1989charged} would produce no such response. A possible detection of 
broken mirror symmetries is significant as it reveals the presence of robust, thermodynamic,
long range ordered phase in the pseudogap regime.

\begin{figure}[t]
\centering \includegraphics[width=0.19\textwidth]{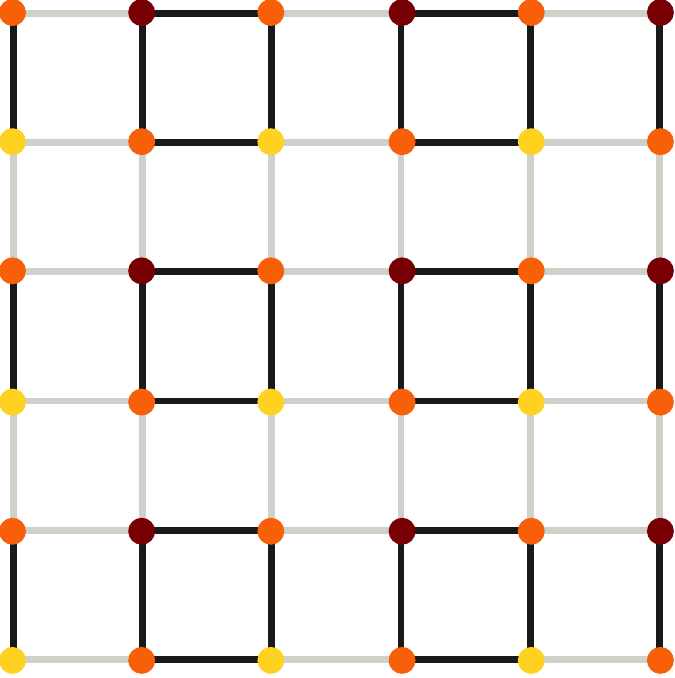}\qquad
\quad \includegraphics[width=0.2\textwidth]{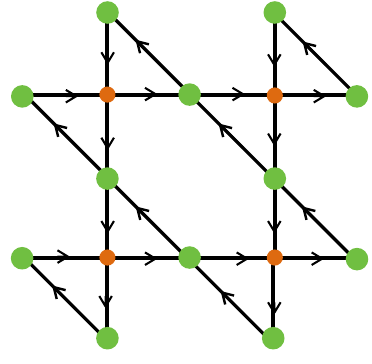}
\caption{\textbf{Left:} A real space picture of a square lattice in the presence of a
CDW and BDW in both $x$ and $y$ directions. Darker points indicate a higher electron density and
darker lines an increased hopping between neighboring sites. \textbf{Right:} A real space picture of the CuO$_2$ plane showing the
$\Theta_{2}$ planar loop order phase which produces a finite shear conductivity. Orange circles represent copper sites, green circles are oxygen sites, and arrows indicate the direction of spontaneously formed microscopic current loops.   }
\label{checkerboardorder} 
\end{figure}

A model calculation can be done by considering the most general Hamiltonian
which includes charge and bond density waves (CDWs and BDWs) in both
$x$- and $y$- directions. This captures the essential mirror-symmetry
breaking physics, and more complicated candidate states (e.g. $d$-wave
charge order) can be constructed from this model. At mean field level,
the density wave Hamiltonian we consider is given by 
\begin{align}
\delta\mathcal{H}_{\text{DW}} & =\sum_{\vec{R}_{i}}\biggl[\phi_{\vec{R}_{i}}c_{\vec{R}_{i}}^{\dagger}c_{\vec{R}_{i}}+\sum_{\alpha=x,y}\bigl(\Delta_{\vec{R}_{i}}^{\alpha}c_{\vec{R}_{i}+\vec{a}_{\alpha}}^{\dagger}c_{\vec{R}_{i}}+\text{H.c.}\bigr)\biggr]
\end{align}
where $\phi_{\vec{R}_{i}}=\phi_{x}\cos(\vec{Q}_{x}\cdot\vec{R}_{i})+\phi_{y}\cos(\vec{Q}_{y}\cdot\vec{R}_{i})$
describes the CDW and $\Delta_{\vec{R}_{i}}^{x/y}=\Delta_{x/y}\cos(\vec{Q}_{x/y}\cdot\vec{R}_{i})$
the BDW. We omit the spin degrees of freedom which will only double the overall
conductivity, and for simplicity we consider period two density waves,
i.e. $|\bm{Q}_{x/y}|=\pi$ . With these choices, a candidate striped
state has (say) finite $\phi_{x}$ and $\Delta_{x}$ but $\phi_{y}=\Delta_{y}=0$,
while a state with both $s$- and $d$-wave charge order in both $x$
and $y$ directions has finite $\phi_{x}=\phi_{y}$, along with finite
$\Delta_{x}=-\Delta_{y}$. With similar generalizations, any candidate
commensurate charge order can be included in this model Hamiltonian,
and while the specific choice of ordering period will affect the following
symmetry considerations, we consider the above an appropriate minimal
model to calculate the shear conductivity.

Before proceeding with a formal calculation, it is useful to anticipate
our results using symmetry arguments. From the real space picture
in Fig.~\ref{checkerboardorder} it is easy to see that the vertical mirror plane
symmetries $\hat{\sigma}_{x}$, and $\hat{\sigma}_{y}$ are broken \textit{only}
when finite period 2 charge \textit{and} bond DW fields in both $x$ and
$y$  directions are present. Finally, the product $\hat{\sigma}_{x=y}*C_{4}$ is also a broken symmetry, since the state is not invariant under 
four-fold rotations. Thus, we only expect
a non-vanishing shear conductance for this system if all charge
and bond fields are finite. Moreover, because 
a mirror reflection is implemented by the change in sign of  {\it any} of the order parameters above, and $\Gamma_{xx,xy}$  much change sign under vertical mirror plane reflections, we expect that the shear conductivity is in general an odd power of each order parameter.  The lowest order contribution would therefore be
linear in each order parameter, \textit{i.e.} $\Gamma\propto\phi_{x}\phi_{y}\Delta_{x}\Delta_{y}$

\begin{figure}[t]
\centering \includegraphics[width=0.45\textwidth]{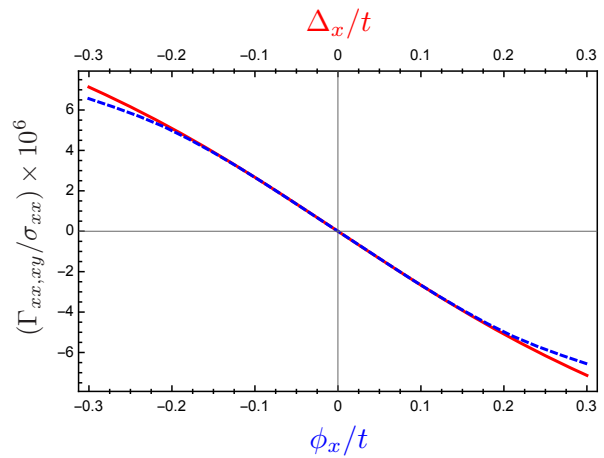}
\caption{Ratio $\Gamma_{xx,xy}/\sigma_{xx}$ of shear conductivity
over longitudinal conductivity for checkerboard charge and bond order
(scattering time $\tau_{\text{sc}}=10/t$). Blue dashed curve as a function
of $\phi_{x}$ for a finite and fixed value of the other three order parameters: $\phi_{y}=\Delta_{x}=-\Delta_{y}=0.1\, t$ and
red curve as a function of $\Delta_{x}$ for similarly fixed $\phi_{x}=\phi_{y}=-\Delta_{y}=0.1\, t$.}
\label{fig:cdwplot}
\end{figure}

To apply this analysis to the cuprate like systems, we use a single
band effective next-nearest neighbor tight-binding dispersion $\epsilon_{\vec{k}}=-2t[\cos(k_{x})+\cos(k_{y})]+4t'\cos(k_{x})\cos(k_{y})-\mu$
of the Cu square lattice with parameters $t'=0.45t$ and $\mu=-0.65t$
(where $t$ is the nearest neighbor hopping for the unstrained system).
The shear strain does not affect the nearest-neighbor hopping at $\mathcal{O}(\epsilon_{xy})$, but does
modify the next nearest neighbor hopping. On the other hand, both the charge and the bond orders are unaffected at this order in strain. Rewriting the Hamiltonian in the four band basis of the new (four site) unit cell, and expanding Eq.~\zref{coe3} to linear order $\epsilon_{xy}$ in the limit
$T\to0$ then leads to $\sigma_{xx}^{\text s}=\sigma_{xx}+\Gamma_{xx,xy}\epsilon_{xy}=P_{x}+2P_{y}\epsilon_{xy}$
with 
\begin{align}
P_{\alpha}=\pi e^{2}\int\limits _{-\pi/2a}^{\pi/2a}\frac{d^{2}\vec{p}}{(2\pi)^{2}}\,\mathbf{Re}\,\text{tr}\biggl[\hat{A}_{\vec{p}}(0)\frac{\partial\hat{H}_{\vec{p}}}{\partial p_{x}}\hat{A}_{\vec{p}}(0)\frac{\partial\hat{H}_{\vec{p}}}{\partial p_{\alpha}}\biggr]\,,\label{cu1}
\end{align}
where the momentum integral is over the new reduced Brillouin zone.
The results of the calculation are shown in Fig.~\ref{fig:cdwplot},
where we plot the ratio $\Gamma_{xx,xy}/\sigma_{xx}$ as a function
of the charge order $\phi_{x}$ and $\Delta_{x}$. Consistent with our expectations based on 
symmetry, the response vanishes when any  of $\phi_x, \phi_y, \Delta_x$, or $\Delta_y$ vanishes, since 
mirror symmetries are partially restored when this happens.   Close to the onset temperature of 
such charge order, we expect the response to be small, as it is proportional to a high power of the order 
parameter fields.  However, there is no reason to expect a small response well below the onset temperature, 
where the order parameters strengths are generically of order unity, in the case of long-range charge order.

\subsection{Shear conductivity as a probe of loop current order}\label{sec:looporder}
While it is now clear that charge ordered states exist within the pseudogap
regime in many cuprates, a more vexing issue is the nature of the
pseudogap itself. The pseudogap onset temperature $T^{*}$ is typically
higher than the charge ordering temperature, with a primary experimental
signature being the suppression of the electronic density of states \cite{NMRYBCO,gurvitch1987resistivity,Ding1996Spectroscopic,timusk1999pseudogap}, and the  onset of $\bm{Q}=0$ magnetism\cite{Bourges2006,Greven2008,Greven2010}. Such magnetic order is consistent with predictions first made by Varma\cite{Varma99Pseudogap,Varma2006}, who argued that various loop current states are responsible for the pseudogap phenomenology.
Recent theoretical work\cite{Tewari2008, aji2013magnetochiral,yakovenko2015tilted,varma2014gyrotropic}
 has suggested that orbital current phases could also explain magneto-optic effects observed in the pseudogap regime\cite{KerrYBCO, Armitage}.
Here we discuss how planar versions of such a phase can be detected
via shear conductivity measurements.

We have considered the so called $\Theta_{1}$ and $\Theta_{2}$ planar
loop current states described in Ref.~\onlinecite{Varma2006}. As Fig.~\ref{checkerboardorder} demonstrates, the $\Theta_{2}$
state breaks all the required symmetries ($\hat{\sigma}_{x},\hat{\sigma}_{y},\hat{\sigma}_{(x=y)}*C_{4}$)
to produce a shear conductivity, while the related $\Theta_{1}$ state with currents on all four diagonal bonds 
preserves the mirror planes $\hat{\sigma}_{x}$ and $\hat{\sigma}_{y}$ and so should not produce
a shear conductivity response.

\begin{figure}[t]
\includegraphics[width=0.45\textwidth]{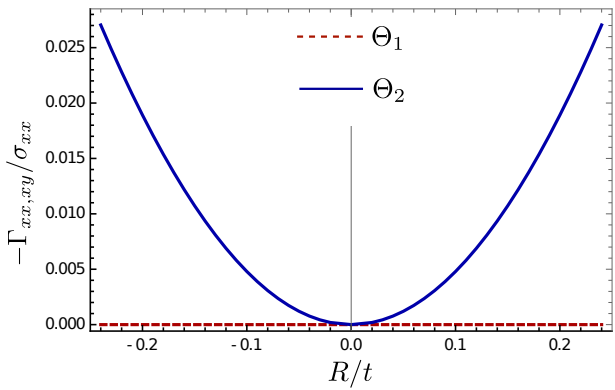} 
\caption{Ratio $\Gamma_{xx,xy}/\sigma_{xx}$ for the loop current
states $\Theta_{1}$ (brown dashes) and $\Theta_{2}$ (blue) as function of
 the order parameter magnitudes $R/t$, where $t$ the tight-binding hopping parameter of the copper-oxygen bond. The response is parametrically larger than for charge order since it is quadratic  in the magnitude of the small, dimensionless order parameters, $R/t$, and not quartic as in the case of charge order.}
\label{Gammaxxxy} 
\end{figure}

For a formal calculation, we work with a three band model that includes
the Cu $d$-orbitals as well as the oxygen $p_{x}$- and $p_{y}$-orbitals
and consider the mean-field Hamiltonian with order parameter $R$ of
the loop current (formally an anapole or toroidal moment, which is a polar vector $\vec{R}$\cite{shekhter2009considerations}). Following the same procedure as before, we implement
strain into this model by modifying the tight binding parameters and
shearing the Brillouin zone. The results are shown in Fig.~\ref{Gammaxxxy}
and, as expected, the system shows a finite shear conductivity $\Gamma_{xx,xy}$
in the symmetry broken state $\Theta_{2}$, but not in $\Theta_{1}$.

We note that unlike the response to charge order, $\Gamma_{xx,xy}$ is even in the current loop order parameter, $\vec{R}$. This follows from the symmetry of the $\Theta_{2}$ state, where the state with order parameter $\vec{R}$ is related to its time reversed partner by a $C_2$ rotation about the $z$ axis (see Fig.~\ref{checkerboardorder}). Because a $C_2 $ rotation leaves the tensor $\Gamma_{xx,xy}$ invariant, we have the equality  $\Gamma_{xx,xy}(\vec{R}) = \Gamma_{xx,xy}(-\vec{R})$. Therefore, the response must be proportional to an even power of $\vec{R}$. 
We stress that the shear conductivity is likely to be considerably larger in the Varma current loop phase than in the charge ordered states studied in the previous section since it is proportional to a smaller power of the order parameter $R$ near the onset temperature, where a Landau-Ginzburg treatment remains valid.

\section{Training the shear conductivity response}\label{sec:training}
Having demonstrated that $\Gamma_{xx,xy}$ is finite when the requisite point group symmetries are broken, it is natural to ask whether the shear conductivity is measurable in macroscopic crystals where domains are inevitably present. Indeed, for both the charge and loop current order parameters, the response from different domains will cancel, as we discuss below. Nevertheless, because the response involves a product of order parameters (\textit{i.e.} it is proportional to a composite order parameter), it is in fact possible to train this composite order parameter (and hence the shear conductivity) across the putative phase transition into a mirror symmetry breaking phase. 

One can anticipate how such training is achieved based on general symmetry considerations: in crystals where a $D_{4h}$ point group symmetry is present in the symmetric phase (the corresponding space group is $P4/mmm)$, $\Gamma_{xx,xy}$ behaves like the $B_{2g}$ ($d_{xy}$) representation of $D_{4h}$, and so must be proportional to a composite order parameter with this symmetry. We can therefore train domains of such a composite order parameter by applying a shear strain, $\epsilon_{xy}$, through a symmetry breaking transition. Below, we discuss the formal symmetry considerations that allow such training for the loop current phase (where the absence of translation symmetry breaking makes the analysis simpler), and then the charge ordered phase.

First, in the case of the $\Theta_{2}$ loop current state, there are four possible domains: two time reversed partners with currents on the $x=y$ diagonals (with order parameters denoted as $\pm\vec{R}^{a}$), and another two with currents on the opposite diagonal ($\pm \vec{R}^{b}$). The responses from domains with currents on opposite diagonals cancel, because they are related by a $C_4$ rotation which sends $\Gamma_{xx,xy}(\vec{R}^{a}) \rightarrow - \Gamma_{yy,yx}(\vec{R}^{b}) = -\Gamma_{xx,xy}(\vec{R}^{b})$, where the second equality follows from $\hat{\sigma}_{(x=y)}$ mirror symmetry of the state (see Appendix~\ref{sec:loopsymm}). Nevertheless, because the responses from time reversed partners on a given diagonal are equal, a macroscopic response is possible if we can bias domains to be of a single orientation. This is easily accomplished by cooling through the loop current transition in the presence of an externally applied $B_{2g}$ strain field (a shear strain $\epsilon_{xy}$), which couples to the orientational component of the loop current order in the free energy like  $\delta F = \lambda \epsilon_{xy}\left(|\vec{R}^{a}|^2 - |\vec{R}^{b}|^2\right)$ where $\lambda$ is a coupling constant.

Next, in the case of the charge order considered in Sec.~\ref{sec:chargeorder}, there are two possible domains of the composite order parameter $\phi_x \phi_y \Delta_x \Delta_y$, which if both present will lead to a vanishing shear conductivity. Nevertheless, while these individual order parameters cannot be trained, their product preserves translation symmetry while only breaking point group symmetries and so can once more be trained by application of an appropriate strain. We can determine the appropriate symmetry channel as follows:
the composite order parameter $\phi_{x}\Delta_{x}$ is translation invariant for the case of $\bm{Q}_{x} = (\pi,0)$. This pattern breaks the $\hat{\sigma}_x$ mirror symmetry, and upon $C_{4}$ rotations, is transformed into the composite order parameter $\phi_{y}\Delta_{y}$. Thus, $\phi_{x}\Delta_{x}$ and $\phi_{y}\Delta_{y}$ have $p_{x}$ and $p_{y}$ symmetry, and so transform like components of the two dimensional $E_{u}$ representation of $D_{4h}$. The product of these two order parameters is therefore in the $B_{2g}$ representation of $D_{4h}$, and there is an allowed coupling to shear strain in the free energy, with a term of the form $\delta F = \lambda \epsilon_{xy}\phi_x \phi_y \Delta_x \Delta_y$. Applying shear strain and cooling through the charge ordering transition should therefore induce a finite shear conductivity response.

\section{Shear resistance measurements}\label{sec:measurements}
Having computed $\Gamma_{xx,xy}$ for various ordered phases, we now 
discuss the experimental protocols required to measure such a response.
Here, we build on recently used experimental methods\cite{Kuo1,Kuo2,Chu12} for determining specific terms in the elastoresistivity
tensor%
\footnote{Note that the same symmetry arguments can be applied to either the
shear conductivity or shear resistivity and moreover $\Gamma_{xx,xy}\neq0\Leftrightarrow m_{xx,xy}\neq0$.%
} 
$m_{\alpha\beta,\gamma\delta}=\frac{\partial\left[\left(\nicefrac{\Delta\rho}{\rho}\right)_{\alpha\beta}\right]}{\partial\epsilon_{\gamma\delta}} \bigr|_{\epsilon\rightarrow0}$ which describes the strain-induced change 
 in the normalized resistivity. The technique, which was originally employed to reveal the presence of nematic fluctuations, can be generalized in the following way. 

\begin{figure}
\centering \includegraphics[width=0.33\textwidth]{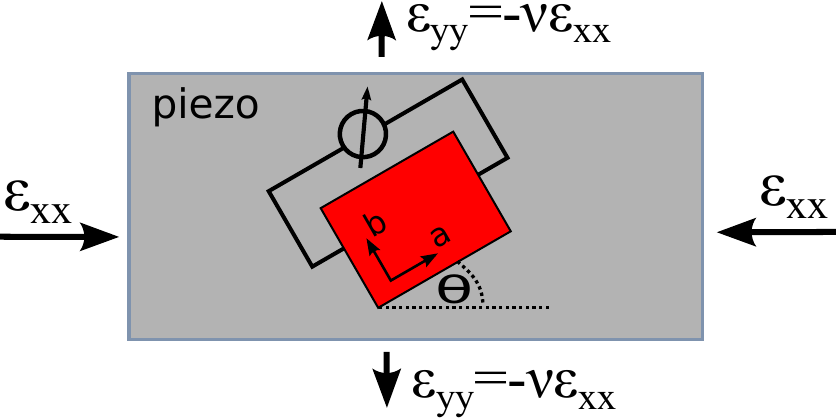} \caption{Experimental setup for the detection of shear resistance similar to
Ref.~{[}\onlinecite{Kuo1,Kuo2}{]}}

\label{ExpSet} 
\end{figure}

A rectangular sample with principal axes $a,b$ oriented along its edges is glued 
with a relative angle $\theta$ onto a piezoelectric
stack in order to measure the longitudinal elastoresistance (Fig.~\ref{ExpSet}).
Applying a voltage to the piezo leads to a stress and therefore tensile
strains $\epsilon_{xx}$ and $\epsilon_{yy}=-\nu_{p}\epsilon_{xx}$
 (where $\nu_{p}$ is the Poisson ratio of the piezoelectric
stack) in the coordinate system of the piezo stack. Since the sample is mounted with a relative angle on the piezo element, the pure tensile strain of the piezo stack translates into
a combination of tensile and shear strain in the coordinate system
of the crystal axes $a,b$ for the sample.

Measuring the change of the resistance $\left(\nicefrac{\Delta R}{R}\right)_{xx}\simeq\left(\nicefrac{\Delta\rho}{\rho}\right)_{xx}$
due to the applied strain for several angles $\theta$, the angularly antisymmetric
component of the response is given by 
\begin{align}
\left(\frac{\Delta R}{R}\right)^{\text{odd}}_{xx}\kern-1em(\theta)&=\frac{1}{2}\left[\left(\frac{\Delta R}{R}\right)_{xx}\kern-1em(\theta)-\left(\frac{\Delta R}{R}\right)_{xx}\kern-1em(-\theta)\right] \\
 & =(\nu_p+1)\,\biggl[\frac{m_{xx,zz}C_{zz,xy}}{C_{zz,zz}}-m_{xx,xy}\biggr]\sin(2\theta)\cdot\epsilon_{xx},\nonumber
\end{align}
where $C_{ij,kl}$ are components of the elastic stiffness tensor. For tetragonal
or orthorhombic symmetry, $C_{zz,xy}=m_{xx,xy}=0$, and the odd contribution
vanishes; conversely, a nonzero  $\left(\nicefrac{\Delta R}{R}\right)^{\text{odd}}_{xx}$
is proof of broken mirror symmetries. An equivalent approach involves
using lithographic methods to prepare samples with different axes
orientations (see Appendix~\ref{expset}).

\section{Discussion}\label{sec:discussion}
We have demonstrated the use of a higher rank
tensor --- the shear conductivity $\Gamma_{xx,xy}$ --- as a sensitive
probe of broken point group symmetries. In the context of the cuprates,
the onset of such a response at the charge ordering temperature would
unambiguously demonstrate that charge order breaks all in-plane mirror
reflection symmetries; recall these are discrete symmetries which
are less susceptible to disorder and can survive as true long range
ordered states. However, it is possible that a current loop state
onsets at a higher, pseudogap temperature scale, in which case the
detection of $\Gamma_{xx,xy}$ at $T^{*}$ supports proposals of the
$\Theta_{2}$ version of this $\bm{Q}=0$ magnetic state. In either
scenario, any identification of broken mirror symmetries is significant
as it specifies a true thermodynamic phase in the pseudogap regime,
and places constraints on a putative quantum critical point which
may exist near optimal doping in cuprate materials.

To summarize, we have considered transport coefficients in the presence of strain that are direct 
indicators of mirror symmetry breaking.  We have studied the shear conductivity in the context of two broken symmetry phases that have been proposed to occur within the pseudogap regime of hole-doped cuprates and have discussed experimental protocols to measure such response.  Our analysis can be generalized to study other components of the strain conductivity tensor $\Gamma_{i j k l} $. 

\section{Acknowledgements}
We thank A.T. Hristov and Chandra Varma for useful discussions. SR would also like to thank the KITP-UCSB Research Program, ÒMagnetism, Bad Metals and SuperconductivityÓ, for hospitality.
PH was supported with a PhD-fellowship of the Deutscher Akademischer Austauschdienst (DAAD). Work at Stanford University was supported by the Department of Energy, Office of Basic Energy Sciences under contract DE-AC02-76SF00515 (AM, PH, MS, SR and IF), and the Alfred P. Sloan Foundation (SR).

\bibliographystyle{h-physrev.bst}
\bibliography{refs}

%---------------------------------------------------------------------------------------------------------------------------------------
%---------------------------------------------------------------------------------------------------------------------------------------

\appendix
\begin{widetext}

\section{Symmetry considerations} \label{secsym} 
\subsection{General properties of tensorial response functions}
Given a Hamiltonian $\mathcal{H}$
that is invariant under the point group transformation $\hat{O}:\vec{r}\rightarrow\vec{r}'$,
 a physical response function $R_{x_{1},\ldots,x_{n}}$, which
is a tensor of rank $n$ (e.g. resistivity, elasticity, \textit{etc.}) is constrained 
according to
\begin{align}
R_{x_{1},\ldots,x_{n}} & =\hat{O}_{x_{1},y_{1}}\ldots\hat{O}_{x_{n},y_{n}}R_{y_{1},\ldots y_{n}}.\label{np1}
\end{align}
This condition typically restricts several elements of $\hat R$ to vanish; conversely,
breaking some point group symmetries relaxes these constraints, allowing the new responses to serve as evidence of broken symmetry. In the following, we will restrict ourselves to the case of (quasi) two-dimensional materials as the various cuprates, iron-based superconductors or heavy-fermion compounds mostly possess a tetragonal symmetry in the high-temperature regime. Nevertheless, it is straightforward to generalize these considerations to more general crystal classes.

Consider now a state that is invariant under reflection about say
the $xy$-plane ($\hat{\sigma}_{z}: \{x,y,z \} \rightarrow \{x,y,-z\}$). We find the condition that
 $$R_{x_{1},\ldots,x_{n}}= (-1)^{N_z} R_{x_{1},\ldots,x_{n}}$$, where $N_z$ is the number of $z$-indices (and likewise for $\hat{\sigma}_{x}$ and $\hat{\sigma}_{y}$ symmetry).  Fourfold rotation symmetry about
the $z$-axis leads to the conditions $$R_{x_{1},\ldots,x_{n}}= (-1)^{N_x} R_{x_{1},\ldots,x_{n}} \bigr|_{x \leftrightarrow y},$$ and  $$R_{x_{1},\ldots,x_{n}}= (-1)^{N_y} R_{x_{1},\ldots,x_{n}} \bigr|_{x \leftrightarrow y}.$$ Finally, a mirror symmetry $\hat{\sigma}_{(x=y)}$ (about the $(x=-y)z$ plane) implies the condition $$R_{x_{1},\ldots,x_{n}}= R_{x_{1},\ldots,x_{n}} \bigr|_{x \leftrightarrow y}.$$

Of particular interest to us is the shear conductivity component $\Gamma_{xx,xy}=\partial\sigma_{xx}/\partial\epsilon_{xy}$  for 2D systems, which describes the change of the longitudinal DC 
conductivity induced by a shear strain of the crystal. Based on the above
symmetry considerations, this can be non-vanishing \textit{only} when reflections
about the $x$- and the $y$-axes -- $\hat{\sigma}_{y}$, $\hat{\sigma}_{x}$ -- as well
as the combination $\hat{\sigma}_{(x=y)}*C_{4}$, where $\hat{\sigma}_{(x=y)}$ denotes reflection
about the vertical $(x=-y)z$ plane and $C_4$ a $\pi/2$ rotation about the $z$-axis, are broken. Following these lines, it is straightforward to construct the explicit forms of higher-rank tensors as the elastoresistance or the elastic stiffness systems with arbitrary symmetries, see e.g., Section~\ref{expset}. 

\subsection{Properties of $\Gamma_{xx,xy}$ for planar loop currents}\label{sec:loopsymm}
We now discuss properties of the shear conductivity within a given loop current state, i.e. a single domain which corresponds to one configuration of the loop current order parameter $\vec{R}$. The magnetic point group symmetry within the loop current state is isomorphic to $D_{2h}$\cite{shekhter2009considerations}, with the following group elements in addition to the identity: $C_{2}(z)\mathcal{T}$, $C_{2}(x=y)\mathcal{T}$, $C_{2}(x=-y)$, $i\mathcal{T}$, $\hat{\sigma}_{z}$, $\hat{\sigma}_{(x=-y)}\mathcal{T}$, and $\hat{\sigma}_{(x=y)}$, where $C_{2}(z)$ denotes a 180 degree rotation about the $z$ axis etc., $\mathcal{T}$ denotes time reversal, $i$ is inversion, and $\hat{\sigma}_{(x=y)}$ denotes mirror reflection as before etc. Note that these are the group elements for the configuration shown in Fig.~\ref{checkerboardorder}. These group operations place the following constraints on the shear conductivity in the presence of a $c$ axis directed magnetic field, $H_z$:
\begin{align}
C_2(z)\mathcal{T},\, i\mathcal{T}:\quad \Gamma_{xx,xy}(\R,H_z) &= \Gamma_{xx,xy}(\R,-H_z) \label{eq:id1}\\
C_2(x=-y),\hat{\sigma}_{(x=y)}:\quad \Gamma_{xx,xy}(\R,H_z) &= \Gamma_{yy,yx}(\R,-H_z)\\
C_2(x=y)\mathcal{T},\hat{\sigma}_{(x=-y)}\mathcal{T}:\quad \Gamma_{xx,xy}(\R,H_z) &= \Gamma_{yy,yx}(\R,H_z).
\end{align}
while $\hat{\sigma}_{z}$ places no constraint. For clarity, note that $C_2\mathcal{T}$ and $i\mathcal{T}$ \textit{independently} imply the identity of Eq.~\ref{eq:id1}, and similarly for the subsequent equations.

\subsection{Properties of $\Gamma_{xx,xy}$ for charge order}
In the case of the charge order described in Sec.~\ref{sec:chargeorder}, the point group symmetry is much more restricted. In the case of equal magnitude charge and bond order parameters as drawn in Fig.~\ref{checkerboardorder}, with $|\phi_{x}| = |\phi_{y}|$ and $|\Delta_{x}| = |\Delta_{y}|$, the point group symmetry is $C_{2v}$, with the group elements $C_{2}(x=-y)$, $\hat{\sigma}_{(x=y)}$, and $\hat{\sigma}_{z}$ in addition to the identity. These enforce the following constraints on the shear conductivity in a $c$ axis directed magnetic field:
\begin{align}
C_2(x=-y), \hat{\sigma}_{(x=y)}:\quad \Gamma_{xx,xy}(H_z) &= \Gamma_{yy,yx}(-H_z) \label{eq:id5}
\end{align}
while $\hat{\sigma}_{z}$ places no constraint. 
Alternatively, when the magnitudes of the order parameters are different, with $|\phi_{x}| \ne |\phi_{y}|$ and $|\Delta_{x}| \ne |\Delta_{y}|$ the point group symmetry is reduced to $C_{s}$, with only two group elements: the identity, and $\hat{\sigma}_{x=y}$. This places no constrains on transformations of $\Gamma_{xx,xy}$.

%---------------------------------------------------------------------------------------------------------------------------------------

\section{Linear Response theory  of elastoconductivity}
\label{sec:linearresponse}
In this section we derive the general expression for the elastoconductivity using Kubo's Linear Response theory. The presented formalism is suitable for any kind of bilinear Hamiltonian with the form
\begin{align}
\mathcal H = \sum_{\vec k, \alpha \beta}   H_{\vec k, \alpha \beta} c_{\vec k,\alpha}^\dagger c_{\vec k,\beta}   = \sum_{\vec k} \hat \Psi_{\vec k}^\dagger \hat H_{\vec k} \hat \Psi_{\vec k},    \label{sa1}
\end{align}
with corresponding Matsubara Green's function $\hat G_{\vec k}(i \omega_n) = \bigl[ i \omega_n \mathds 1 - \hat H_{\vec k}\bigr]^{-1}$. Following Mahan~\cite{mahan2000many} the longitudinal DC conductivity for such a multiband system can be calculated to be
\begin{align}
\sigma_{xx} = - \pi e^2 \int \limits_{\text{1st B.Z.}} \frac{d^d \vec k}{(2\pi)^d}  \int \limits_{-\infty}^{\infty} d\epsilon \frac{\partial f(\epsilon)}{\partial \epsilon} \mathbf{Re} \, \text{tr} \biggl[ \bigl( \hat A_{\vec k}(\epsilon) \frac{\partial \hat H_{\vec k} }{\partial k_x} \bigr)^2 \biggr],    \label{sa2}
\end{align}
where $\hat A_{\vec k}(\epsilon) = \frac{i}{2\pi} \bigl[ \hat G_{\vec k}(\epsilon+i / 2\tau_{\text{sc}}) - \hat G_{\vec k}(\epsilon-i / 2\tau_{\text{sc}}) \bigr]$ is the spectral weight of the system, $\epsilon$ is an energy and $f(\epsilon)$ is the usual Fermi-Dirac distribution function. In order to yield a finite result of this transport expression, it is necessary to introduce a finite lifetime $\tau_{\text{sc}}$ for the fermionic quasiparticles, which can originate from electron-electron, electron-phonon or electron-impurity interactions. Here, we just assume $\tau_{\text{sc}} \neq 0$, and neglect other processes (e.g., vertex corrections). Thus, the scattering rate $1/ \tau_{\text{sc}} \ll E_{F}$ (where $E_{F}$ is the Fermi energy) can be viewed simply as an effective parameter of the theory. 

We now focus on two spatial dimensions and consider a system with a Hamiltonian~\zref{sa1}, to be described by a tight-binding model
\begin{align}
\mathcal H &= \sum_{\vec R_i} \sum_{\alpha,\beta} \epsilon^{\alpha \beta} c_{i,\alpha}^\dagger c_{i,\beta} +\sum_{ <\vec R_i, \vec R_j>} \sum_{\alpha,\beta} \bigl[ t_{ij}^{\alpha \beta} c_{i,\alpha}^\dagger c_{j,\beta} + \text{h.c.}   \bigr]   +\sum_{ <<\vec R_i, \vec R_j>>} \sum_{\alpha,\beta} \bigl[ (t')_{ij}^{\alpha \beta} c_{i,\alpha}^\dagger c_{j,\beta} + \text{h.c.}   \bigr] + \ldots, \label{sa3}
\end{align}
where $\vec R_i$ are the positions of the atoms of the lattice (Bravais vectors $\vec a_1, \vec a_2$) and $\alpha$ describes for example possible orbital or spin degrees of freedom. The matrix/hopping elements $(t^n)_{ij}^{\alpha \beta}$ are in general dependent on the distance between the atoms, which can be parameterized as
\begin{align}
(t^{n})_{ij}^{\alpha,\beta} =\bigr[t^n  (\abs{\vec R_i-\vec R_j} )\bigr]^{\alpha \beta}  = (t^n)^{\alpha \beta} e^{- \gamma_n^{\alpha \beta} \bigl[\abs{\vec R_i-\vec R_j}-\abs{\vec R_i^0-\vec R_j^0}\bigr]},   \label{sa3b}
\end{align}
where $\vec R_i^0$ are the equilibrium positions of the atoms. Without strain, the Fourier space representation of \zref{sa3} is just given by
\begin{align}
\mathcal H &= \sum_{\vec k} \sum_{\alpha,\beta}   c_{\vec k,\alpha}^\dagger c_{\vec k,\beta}  \biggl[  \epsilon^{\alpha \beta} + 2 t^{\alpha \beta} \bigl[\cos(\vec a_1 \cdot \vec k)  + \cos(\vec a_2 \cdot\vec k)    \bigr]    \nonumber \\
& \hspace{0.3cm}+2 (t')^{\alpha \beta} \bigl[\cos( [\vec a_1+\vec a_2] \cdot\vec k)  + \cos([\vec a_1-\vec a_2] \cdot\vec k)    \bigr] + \ldots    \biggr]     \nonumber \\
&= \sum_{\vec k} \Psi_{\vec k}^\dagger \hat H_{\vec k} \Psi_{\vec k}, \label{sa3c}
\end{align}
with the spinor $\Psi_k=(c_{k,1} , c_{k,2} , \ldots )^T$ describing the multiband system. We now apply a general strain to the crystal, which has the form
\begin{align}
\hat \epsilon = \mx{\epsilon_{xx} & \epsilon_{xy} \\ \epsilon_{xy} & \epsilon_{yy}} .  \label{sa4}
\end{align}
The strain changes the Bravais lattice vectors $\vec a_1, \vec a_2$ of the system to
\begin{align}
\vec a_{j}\s = (\hat{ \mathds 1}+ \hat \epsilon) \vec a_{j},\quad j=1,2   \label{sa5}
\end{align}
and therefore also the positions of the atoms $\vec R_i^0 \rightarrow \vec R_i\s = (\hat{ \mathds 1}+ \hat \epsilon) \vec R_i$. The Hamiltonian in the presence of strain is therefore given by
\begin{align}
\mathcal H\s &= \sum_{\vec k}{}^{\prime} \sum_{\alpha,\beta}   c_{\vec k,\alpha}^\dagger c_{\vec k,\beta}  \biggl[  \epsilon^{\alpha \beta} + 2 t^{\alpha \beta} \sum_{j=1,2} e^{- \gamma^{\alpha \beta} \bigl[\abs{ (\hat{ \mathds 1}+ \hat \epsilon) \vec a_{j}}-\abs{\vec a_j}\bigr]} \cos([ (\hat{ \mathds 1}+ \hat \epsilon) \vec a_{j}]\cdot\vec k)  \nonumber \\
& \hspace{2.7cm}+ 2 (t')^{\alpha \beta} \sum_{j=\pm } e^{- \gamma'^{\alpha \beta} \bigl[\abs{ (\hat{ \mathds 1}+ \hat \epsilon) (\vec a_1 +j \vec a_2)}-\abs{\vec a_1 +j \vec a_2}\bigr]} \cos([(\hat{ \mathds 1}+ \hat \epsilon)(\vec a_1 +j \vec a_2)] \cdot \vec k)    + \ldots   \biggr]    \nonumber \\
&=  \sum_{\vec k}{}^{\prime}   \,  \Psi_{\vec k}^\dagger \hat H_{(\hat{ \mathds 1}+ \hat \epsilon) \vec k}\s \Psi_{\vec k}. \label{sa6}
\end{align}

Note that there are now three major changes if we compare \zref{sa6} with \zref{sa3c}: i) Due to \zref{sa5}, the Brillouin zone is strained, which is indicated by $ \sum_{\vec k}{}^{\prime}$; ii) the Hamiltonian matrix 
\begin{align}
 \hat H_{(\hat{ \mathds 1}+ \hat \epsilon) \vec k}\s = \hat H_{(\hat{ \mathds 1}+ \hat \epsilon) \vec k}  \biggr|_{(t^n)_{ij}^{\alpha \beta} \rightarrow (t^n)_{ij}^{\alpha \beta}(\hat \epsilon) },   \label{sa7}
 \end{align} 
has modified tight-binding parameters depending on the strain $\hat \epsilon$; and iii) the quasimomentum $\vec k \rightarrow (\hat{ \mathds 1}+ \hat \epsilon) \vec k$ takes into account the change of the Brillouin zone.  Inserting now our strained Hamiltonian \zref{sa6} into the linear response theory we find
\begin{align}
\sigma_{xx}\s &= - \pi e^2 \int \limits_{\text{strained} \atop \text{1st B.Z.}} \frac{d^2 \vec k}{(2\pi)^2}  \int \limits_{-\infty}^{\infty} d\epsilon \frac{\partial f(\epsilon)}{\partial \epsilon} \mathbf{Re} \, \text{tr} \biggl[ \biggl( \hat A_{(\hat{ \mathds 1}+ \hat \epsilon) \vec k}\s(\epsilon) \frac{\partial \hat H_{(\hat{ \mathds 1}+ \hat \epsilon) \vec k}\s }{\partial k_x} \biggr)^2 \biggr]     \nonumber \\
&=- \frac{\pi e^2}{\text{det}(\hat{ \mathds 1}+ \hat \epsilon)} \int \limits_{ \text{1st B.Z.}} \frac{d^2 \vec p}{(2\pi)^2}  \int \limits_{-\infty}^{\infty} d\epsilon \frac{\partial f(\epsilon)}{\partial \epsilon} \mathbf{Re} \, \text{tr} \biggl[ \biggl( \hat A_{\vec p}\s(\epsilon)  \bigl[ (1+\epsilon_{xx}) \frac{\partial \hat H_{\vec p}^s }{\partial p_x} + \epsilon_{xy}  \frac{\partial \hat H_{\vec p}^s }{\partial p_y} \bigr]  \biggr)^2 \biggr] ,   \label{sa8}
\end{align}
where in the second line we substituted $\vec p= (\hat{ \mathds 1}+ \hat \epsilon) \vec k$ so that the integral over the original (unstrained) Brillouin zone is recovered. In the following we will always differentiate between the momentum $\vec k$ defined in the strained Brillouin zone and $\vec p$ in the unstrained Brillouin zone (see Fig.~\ref{BZ}).

 \begin{figure}
\centering
\includegraphics[width=0.4\textwidth]{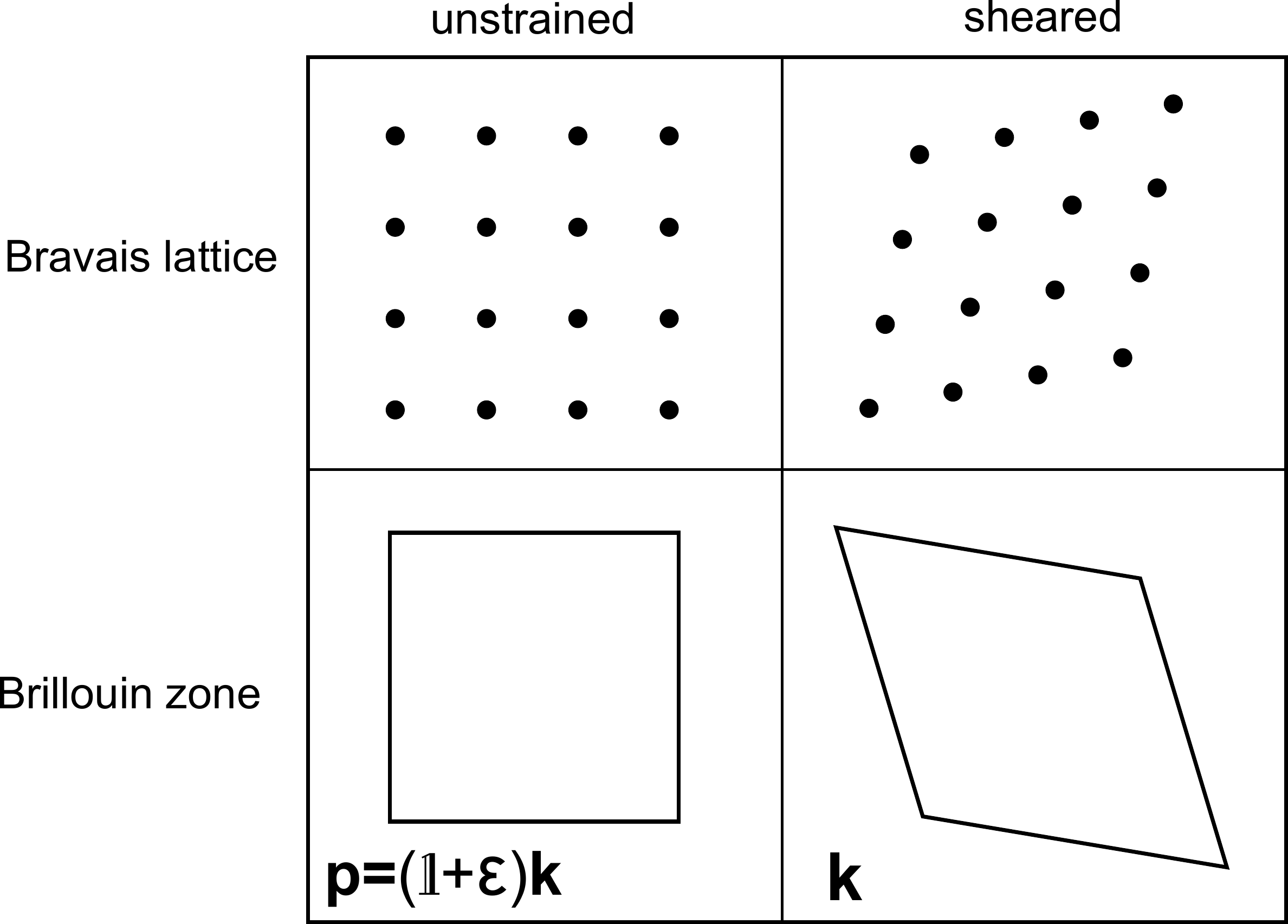}
\caption{Unstrained and sheared Bravais lattice and corresponding first Brillouin zone of square lattices.}
\label{BZ}
\end{figure}

\subsection{Next-nearest neighbor tight-binding dispersion for square lattice in the presence of shear strain}   \label{nnn}
In order to make the connection between the generalized tight-binding Hamiltonian in \zref{sa3c} to the systems considered in this paper, we now calculate the dispersion of a simple square lattice in next-nearest neighbor approximation and in the presence of a pure shear strain
\begin{align}
\hat \epsilon = \mx{ 0 & \epsilon_{xy} \\ \epsilon_{xy} & 0}  \, . \label{sa10}
\end{align}
The Bravais vectors of a square lattice are $\vec a_1=\vec e_x, \vec a_2 = \vec e_y$, where we set the lattice constant $a=1$. The nearest-neighbor hoppings $t\s = t$ (and also the charge and bond order parameters) do not change linearly in $\epsilon_{xy}$ by a pure shear strain since the distance between neighbor atoms is $\abs{ (\hat{ \mathds 1}+ \hat \epsilon) \vec e_{x/y}} = 1 +\frac{1}{2} \epsilon_{xy}^2 \approx 1$. In contrast, the next-nearest neighbor hoppings 
\begin{align}
 \bigl[t' (\vec e_x\s \pm \vec e_y\s) \bigr]\s = t' e^{- \gamma' \bigl( \abs{\vec e_x\s \pm \vec e_y\s}-\abs{\vec e_x \pm \vec e_y}    \bigr) } \approx t' \bigl( 1 \mp \sqrt{2} \gamma' \epsilon_{xy}  \bigr),   \label{sa11}
\end{align}
obtain a linear $\epsilon_{xy}$-dependence and  we get the sheared dispersion 
\begin{align}
 \epsilon_{\vec p}\s &= 2 t \bigl[ \cos(p_x) +\cos(p_y)    \bigr]  +2 t'  \bigl( 1 - \sqrt{2} \gamma' \epsilon_{xy}  \bigr) \cos( p_x+p_y)   \nonumber \\
 &  \qquad +2 t'  \bigl( 1 +\sqrt{2} \gamma' \epsilon_{xy}  \bigr) \cos( p_x-p_y)    - \mu.  \label{sa12}
\end{align}
Note that $\vec p$ is, as mentioned in the last section, the momentum in the unstrained Brillouin zone which has to be used in formula \zref{sa8}.

%---------------------------------------------------------------------------------------------------------------------------------------

\section{Shear conductivity in cuprates with charge order}
\subsection{Effective Hamiltonian for bidirectional charge order}
A bidirectional charge ($\phi$) and bond  ($\Delta$) density wave of period $N \in \mathbb N$ on a square lattice (which describes our quasi two-dimensional material) is described by the commensurate fields
\begin{align}
\begin{split}
\phi_{\vec R_i} &= \phi_{x} \cdot  \cos(\vec Q_{x} \cdot \vec R_i)+ \phi_{y} \cdot  \cos(\vec Q_{y} \cdot \vec R_i),    \\
\Delta^{x/y}_{\vec R_i} &= \Delta_{x/y} \cdot  \cos(\vec Q_{x/y} \cdot \vec R_i),
\end{split}   \label{cu1}
\end{align}
where $\phi_{x/y}, \Delta_{x/y} \in \mathbb{R}$ and $\vec Q_{x/y} = \frac{2\pi}{N} \vec e_{x/y}$. Note that $C_4$ symmetric (i.e., ``checkerboard") order with $\phi_x=\phi_y$ and $\Delta_x=\Delta_y$, also preserves the $\hat{\sigma}_{(x=-y)}$ symmetry. We consider here the effective square copper lattice (Bravais lattice $\vec a_x=\vec e_x, \vec a_y=\vec e_y$) describing the material's dispersion (note that we have ignored spin degrees of freedom which simply contribute an overall factor of 2 in the conductivity). The effective Hamiltonian reads
\begin{align}
\mathcal H = \sum_{\vec R_i} \bigl[\phi_{\vec R_i} - \mu\bigr] c_i^\dagger c_i +\biggl[ - \sum_{\alpha=x,y} \sum_{\vec R_j=\vec R_i+\vec a_\alpha} \bigl[ t+ \Delta^\alpha_{\vec R_i} \bigr] c_j^\dagger c_i + t' \sum_{\vec R_j = \vec R_i + \vec a_x \pm \vec a_y } c_j^\dagger c_i + \text{h.c.} \biggr],   \label{cu2}
\end{align}
where we use the parameters $t=1, t'=0.45 t, \mu=-0.65 t$. Transforming this Hamiltonian into Fourier space leads to
\begin{align}
\mathcal H &= \sum_{\vec k} \epsilon_{\vec k} c_{\vec k}^\dagger c_{\vec k} + \sum_{\vec k} \sum_{\alpha=x,y} \biggl(  \biggl[ \frac{\phi_\alpha}{2} - \Delta_\alpha e^{-i \vec Q_\alpha \cdot \vec a_\alpha} \cos\bigl[ (\vec k+ \frac{\vec Q_\alpha}{2} ) \cdot \vec a_\alpha \bigr] \biggr] c_{\vec k + \vec Q_\alpha}^\dagger c_{\vec k}+ \text{h.c.}   \biggr),   \label{cu3}
\end{align}
with the free dispersion
\begin{align}
\epsilon_{\vec k} = -2 t [\cos(k_x)+\cos(k_y)] + 4 t' \cos(k_x) \cos(k_y) -\mu.    \label{cu4}
\end{align}
Due to the density waves the periodicity of the crystal changes from $a=1$ to $N$, so we have downfolded our original Brillouin zone
 \begin{align}
k_{x/y} \in [-\pi,\pi]    \quad \rightarrow \quad k_{x/y} \in [-\pi/N,\pi/N],   \label{cu5}  
 \end{align}
so that we end up with an effective band Hamiltonian
\begin{align}
\mathcal H &= \sum_{\vec k}  \Psi_{\vec k}^\dagger \hat H_{\vec k} \Psi_{\vec k},   \label{cu6}
\end{align}
where $\Psi_{\vec k} =  (c_{\vec k} , c_{\vec k+\vec Q_x} , \ldots )^T$. From this point on we will restrict ourselves to the case of $N=2$, but the following arguments and steps are valid for arbitrary $N$. For $N=2$ we have the spinor $\Psi_{\vec k} =  (c_{\vec k} , c_{\vec k+\vec Q_x} ,c_{\vec k+\vec Q_y},c_{\vec k+\vec Q_x+\vec Q_y} )^T$ and the Hamiltonian
\begin{align}
\hat H_{\vec k} &= \mx{ \epsilon_{\vec k} & \chi_x(\vec k) & \chi_y(\vec k) & 0 \\
 \chi_x^*(\vec k) & \epsilon_{\vec k+\vec Q_x} & 0 & \chi_{y}(\vec k+\vec Q_x) \\
\chi_y^*(\vec k) & 0 &\epsilon_{\vec k+\vec Q_y} & \chi_x(\vec k + \vec Q_y) \\
0 &\chi_y^*(\vec k+\vec Q_x) & \chi_x^*(\vec k + \vec Q_y)&\epsilon_{\vec k+\vec Q_x+\vec Q_y} 
}\label{cu7},
\end{align}
with
\begin{align}
\chi_{x/y}(\vec k) = \phi_{x/y} +2 \Delta_{x/y}  e^{-i \vec Q_{x/y} \cdot \vec a_{x/y} }\cos\bigl[ (\vec k + \vec Q_{x/y}/2) \cdot \vec a_{x/y} \bigr]  = \phi_{x/y} -2i \Delta_{x/y}  \cos\bigl[ k_{x/y} + \pi/2 \bigr]. \label{cu8}
 \end{align}
 In Fig.~\ref{df} we show the original cuprate Fermi surface and the down-folded Fermi surface for the period $2$ charge and bond density waves.
 
\begin{figure}
 \centering
 \includegraphics[width=0.35\textwidth]{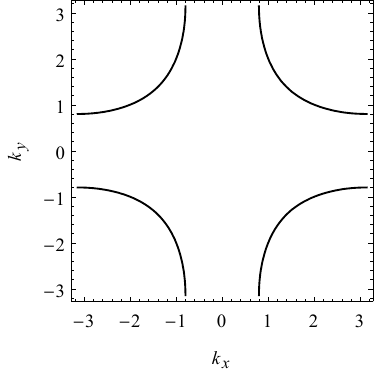}  \includegraphics[width=0.35\textwidth]{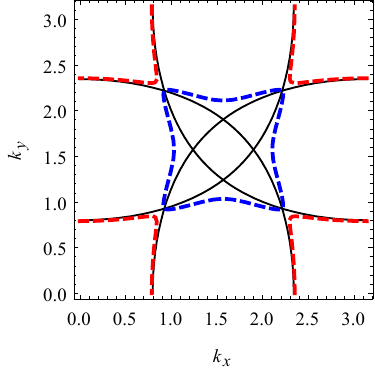}
 \caption{Fermi surface of the original (left) and down-folded (right) Brillouin zone for charge and bond order ($N=2$) with gaps opening at the hot spots of the Fermi surface that are connected by $\vec Q_{x/y} = \pi \vec e_{x/y}$. Note that the figure on the right is the reduced Brillouin zone shown here as the upper right corner of the original Brillouin zone.}
 \label{df}
 \end{figure}

\subsection{Shear conductivity}
We now discuss how to apply the shear strain \zref{sa10} on the density wave system. The change of the Bravais-lattice $\vec a_{x/y}\s = (\mathds 1+ \hat \epsilon) \vec a_{x/y}$ will also lead to a change of the reciprocal lattice vectors $\vec b_{x/y}\s = \frac{2\pi}{V} (\vec e_z \times \vec a_{y/x}\s)$ and finally also the ordering vector $\vec Q_{x/y}\s = \frac{1}{N} \vec b_{x/y}\s$. This means that
\begin{align}
\vec Q_{\alpha} \cdot \vec a_\beta = \vec Q_{\alpha}\s \cdot \vec a_\beta \s = \frac{2\pi}{N},   \label{cu9}
\end{align}
remains invariant under the application of the strain. As we saw in Section~\ref{nnn}, the nearest neighbor hoppings are not affected linearly in $\epsilon_{xy}$ by a pure shear strain and the same argument should therefore hold for the charge and bond order $\phi_{x/y}, \Delta_{x/y}$ since their ordering vectors are aligned with the $x$ and $y$ axis. This means that the strained Hamiltonian of \zref{cu7} is given by
\begin{align}
\hat H\s_{\vec p} &= \mx{ \epsilon\s_{\vec p} & \chi_x(\vec p) & \chi_y(\vec p) & 0 \\
 \chi_x^*(\vec p) & \epsilon\s_{\vec p+\vec Q_x} & 0 & \chi_{y}(\vec p+\vec Q_x) \\
\chi_y^*(\vec p) & 0 &\epsilon\s_{\vec p+\vec Q_y} & \chi_x(\vec p + \vec Q_y) \\
0 &\chi_y^*(\vec p+\vec Q_x) & \chi_x^*(\vec p + \vec Q_y)&\epsilon\s_{\vec p+\vec Q_x+\vec Q_y},
}\label{cu10}
\end{align}
 with the strained dispersion $\epsilon\s_{\vec p}$ defined in \zref{sa12}. We now rewrite the formula for the elastoconductivity in the case of  pure shear strain 
\begin{align}
\sigma_{xx}\s &\verweis{sa8} - \frac{\pi e^2}{1-\epsilon_{xy}^2} \underbrace{\int \limits_{ \text{1st B.Z.}}}_{\int_{-\pi/2}^{\pi/2}} \frac{d^2 \vec p}{(2\pi)^2}  \int \limits_{-\infty}^{\infty} d\epsilon \frac{\partial f(\epsilon)}{\partial \epsilon} \mathbf{Re} \, \text{tr} \biggl[ \bigl( \hat A_{\vec p}\s(\epsilon) \hat J_{\vec p,x}\s  \bigr)^2 \biggr],     \label{cu11}
\end{align}
where the integral is performed over the reduced Brillouin zone and we have defined the current vertex as
\begin{align}
\hat J_{\vec p,x}\s = \frac{\partial \hat H_{\vec p}^s }{\partial p_x} + \epsilon_{xy}  \frac{\partial \hat H_{\vec p}^s }{\partial p_y}.    \label{cu12}
\end{align}
Expanding the spectral weight and the current operator to linear order in the shear strain $\epsilon_{xy}$ we have
\begin{align}
\begin{split}
 \hat A_{\vec p}\s(\epsilon) &=  \hat A_{\vec p}\s(\epsilon) \biggr|_{\epsilon_{xy}=0} +  \frac{\hat A_{\vec p}\s(\epsilon)}{\partial \epsilon_{xy}} \biggr|_{\epsilon_{xy}=0} \cdot \epsilon_{xy}  =  \hat A_{\vec p}(\epsilon)  + \delta \hat A_{\vec p}(\epsilon) \cdot \epsilon_{xy}   \\
  \hat J_{\vec p,x}\s&=  \hat J_{\vec p,x}\s \biggr|_{\epsilon_{xy}=0} +  \frac{\hat J_{\vec p,x}\s}{\partial \epsilon_{xy}} \biggr|_{\epsilon_{xy}=0} \cdot \epsilon_{xy}  =   \hat J_{\vec p,x}  + \delta    \hat J_{\vec p,x}\cdot \epsilon_{xy}.
  \end{split}  \label{cu13}
\end{align}
Inserting this in \zref{cu11} and expanding again, we find the DC conductivity
\begin{align}
\sigma_{xx} &= - \pi e^2 \int_{-\pi/2}^{\pi/2} \frac{d^2 \vec p}{(2\pi)^2}  \int \limits_{-\infty}^{\infty} d\epsilon \frac{\partial f(\epsilon)}{\partial \epsilon} \mathbf{Re} \, \text{tr} \biggl[ \bigl( \hat A_{\vec p}(\epsilon) \hat J_{\vec p,x}  \bigr)^2 \biggr],  \label{cu14}
\end{align}
and the shear conductivity 
\begin{align}
\Gamma_{xx,xy} &= - 2\pi e^2 \int_{-\pi/2}^{\pi/2} \frac{d^2 \vec p}{(2\pi)^2}  \int \limits_{-\infty}^{\infty} d\epsilon \frac{\partial f(\epsilon)}{\partial \epsilon} \mathbf{Re} \, \text{tr} \biggl[  \delta \hat A_{\vec p}(\epsilon) \hat J_{\vec p,x}  \hat A_{\vec p}(\epsilon) \hat J_{\vec p,x} + \hat A_{\vec p}(\epsilon) \delta \hat J_{\vec p,x}  \hat A_{\vec p}(\epsilon) \hat J_{\vec p,x} \biggr].    \label{cu15}
\end{align}
A numerical investigation shows that $\Gamma_{xx,xy}$ is insensitive to $\gamma'$, so we henceforth set $\gamma'=0$. In this case we can finally simplify
\begin{align}
\sigma_{xx} = Q_x,   \hspace{2cm} \Gamma_{xx,xy}=2 Q_y,    \label{lin12}
\end{align}
with
\begin{align}
Q_{\alpha} &=  - \pi e^2 \int_{-\pi/2}^{\pi/2} \frac{d^2 \vec p}{(2\pi)^2}  \int \limits_{-\infty}^{\infty} d\epsilon \frac{\partial f(\epsilon)}{\partial \epsilon} \mathbf{Re} \, \text{tr} \biggl[  \hat A_{\vec p}(\epsilon) \frac{\partial \hat H_{\vec p}}{\partial p_x}  \hat A_{\vec p}(\epsilon) \frac{\partial \hat H_{\vec p}}{\partial p_\alpha} \biggr]   \nonumber \\
&\xrightarrow{T \rightarrow 0} \pi e^2 \int_{-\pi/2}^{\pi/2} \frac{d^2 \vec p}{(2\pi)^2}  \mathbf{Re} \, \text{tr} \biggl[  \hat A_{\vec p}(0) \frac{\partial \hat H_{\vec p}}{\partial p_x}  \hat A_{\vec p}(0) \frac{\partial \hat H_{\vec p}}{\partial p_\alpha} \biggr],    \label{lin13}
\end{align}
where in the end we considered the zero temperature limit. This formula is then numerically evaluated to find the shear conductivity numerically, as discussed in the main text.

%---------------------------------------------------------------------------------------------------------------------------------------

\section{Shear conductivity in cuprates for  current loop order}
Following Ref.~[\onlinecite{Varma2006}] we calculate the shear conductivity for the Varma current loop states $\Theta_I$ and $\Theta_{II}$. From the minimal model of the copper oxides that takes into account the $d_{x^2-y^2}$ orbitals of the copper atoms and the $p_{x},p_y$ orbitals of the oxygen atoms we can write down the mean-field Hamiltonian of the time-reversal breaking fields as (spin degrees of freedom are introduced later)
\begin{align}
\begin{split}
\mathcal H_{\Theta_I} &= \sum_{\vec k} \mx{d_{\vec k} \\ p_{x,\vec k} \\ p_{y,\vec k}}^\dagger \overbrace{\mx{ \epsilon_d - \mu & [2 t_{pd}- R e^{i \phi}] s_x(\vec k) & [2 t_{pd} + R e^{i \phi}] s_y(\vec k) \\ \text{h.c.} & - \mu & - 4 t_{pp} s_x(\vec k) s_y(\vec k) \\ \text{h.c.} & \text{h.c.}&  - \mu  }}^{\hat H_{\vec k}^{\Theta_I}}\mx{d_{\vec k} \\ p_{x,\vec k} \\ p_{y,\vec k}}, \\
\mathcal H_{\Theta_{II}} &= \sum_{\vec k} \mx{d_{\vec k} \\ p_{x,\vec k} \\ p_{y,\vec k}}^\dagger \underbrace{\mx{ \epsilon_d - \mu & 2 t_{pd}  s_x(\vec k)- R e^{i \phi} c_x(\vec k) & 2 t_{pd} s_y(\vec k) + R e^{i \phi}  c_y(\vec k)\\ \text{h.c.} & - \mu & - 4 t_{pp} s_x(\vec k) s_y(\vec k) \\ \text{h.c.} & \text{h.c.}&  - \mu  } }_{\hat H_{\vec k}^{\Theta_{II}}} \mx{d_{\vec k} \\ p_{x,\vec k} \\ p_{y,\vec k}}, 
\end{split}
\end{align}
where $t_{pd}$ and $t_{pp}$ are the next nearest neighbor hopping parameters of the copper-oxygen and oxygen-oxygen bonds, $\epsilon_d$ the on-site energy of the copper orbitals, $\mu$ the chemical potential, $R \cdot e^{i \phi}$ the current loop order parameter of the Varma theory (with $\phi=\pm \pi/2$) and we defined $s_{x/y}(\vec k) = \sin(\vec k \cdot \vec a_{x/y})$, $c_{x/y}(\vec k) = \cos(\vec k \cdot \vec a_{x/y})$ with $\vec a_{x/y}= a \cdot \hat e_{x/y}$. The creation/annihilation operators are defined in such a way that the current vertex is again just given by 
\begin{align}
\vec J_{\vec k}^{\Theta_{I/II}} =  \frac{\partial \hat H_{\vec k}^{\Theta_{I/II}}}{\partial \vec k}.
\end{align}
Introducing the shear as in the previous section and assuming a linear shear-dependence of $t_{pp}^{\nearrow}(\epsilon_{xy})=t_{pp}(1-\gamma  \epsilon_{xy})$ and $t_{pp}^{\searrow}(\epsilon_{xy})=t_{pp}(1+\gamma  \epsilon_{xy})$  (where arrows schematically indicate the direction of oxygen-oxygen bonds on the diagonal), we can  expand expression~\zref{sa8} linear in strain. For the numerical investigation we use the  parameters $t_{pp}= 0.3 t_{pd}, \epsilon_d = - 0.7 t_{pd}, \mu=1.5 t_{pd}, \tau = 10/t_{pd}, \gamma=1.3 t_{pd} $. As discussed in the main text, the $\Theta_{I}$ state, which does not break the $\hat{\sigma}_{z}$ and $\hat{\sigma}_{y}$ mirror symmetry, does not produce a finite shear conductivity --- $\Gamma_{xx,xy} = 0$. In contrast, the $\Theta_{II}$ state breaks the requisite mirror symmetries and hence shows a finite shear conductivity $\Gamma_{xx,xy} \neq 0$ which increases with the order parameter $R$. Note that it is the $\Theta_{II}$ state (and several symmetry related partner states involving out of plane current loops to apical oxygens) which is both a stronger theoretical and experimental candidate for a pseudogap state.

%---------------------------------------------------------------------------------------------------------------------------------------

\section{Experimental Setups}   \label{expset}
\subsection{Explicit forms of tensors for broken tetragonal symmetry}
Let us now consider (quasi) two-dimensional materials with an intrinsic $\hat{\sigma}_{z}$ symmetry, e.g., the layered cuprate or iron pnictide compounds. If all tetragonal point group symmetries $\hat{\sigma}_{x}, \hat{\sigma}_{y}, \hat{\sigma}_{(x=-y)}, C_4$ are broken, a generic rank-4 tensor only has components with an even number of $z$-indices, see Sec.~\ref{secsym}. In the usual Voigt notation $(1,2,3,4,5,6)=(xx,yy,zz,yz,zx,xy)$ we can then write a rank 4 tensor $\hat T$ that is symmetric in the first and last two indices (such as the elastoresistance or elastic stiffness)\footnote{The elastoresistance $m_{\alpha \beta,\gamma \delta} = \partial R_{\alpha \beta}/\partial \epsilon_{\gamma \delta}$ is symmetric in the first two indices due to Onsager's relation $R_{\alpha \beta}(\vec H)= R_{\beta \alpha}(\vec H)$ (in the absence of a magnetic field $\vec H$) and in the last two indices due to the symmetric definition of the strain tensor $\epsilon_{\gamma \delta}= \epsilon_{\delta \gamma}$.} as
 \begin{align}
 \text{Symmetry: }\hat{\sigma}_{z} \hspace{2cm} \hat T =\mx{
 T_{11} & T_{12} & T_{13} & 0 & 0 & T_{16} \\
 T_{21} & T_{22} & T_{23} & 0 & 0 & T_{26} \\
 T_{31} & T_{32} & T_{33} & 0 & 0 & T_{36} \\
 0 & 0 & 0 & T_{44} & T_{45} & 0 \\
 0 & 0 & 0 & T_{54} & T_{55} & 0 \\
 T_{61} & T_{62} & T_{63} & 0 & 0 & T_{66} }. \label{bts1}
 \end{align}
Let us now consider what happens if we restore the different point group symmetries $\hat{\sigma}_{y}, \hat{\sigma}_{z}, \hat{\sigma}_{(x=-y)}, C_4$ of the tetragonal group. Restoring either $\hat{\sigma}_x$ or $\hat{\sigma}_y$ will lead to the vanishing of all tensor elements with an odd number of $x$-indices (and therefore also for the elements with an odd number of $y$-indices):
\begin{align}
 \text{Symmetries: }\hat{\sigma}_{z}, (\hat{\sigma}_{x}  \text{ or }  \hat{\sigma}_{y}) \hspace{2cm} \hat T =\mx{
 T_{11} & T_{12} & T_{13} & 0 & 0 & 0 \\
 T_{21} & T_{22} & T_{23} & 0 & 0 &0 \\
 T_{31} & T_{32} & T_{33} & 0 & 0 & 0 \\
 0 & 0 & 0 & T_{44} & 0 & 0 \\
 0 & 0 & 0 & 0 & T_{55} & 0 \\
 0 & 0& 0 & 0 & 0 & T_{66} } \label{bts2}
 \end{align}
Restoring the $\hat{\sigma}_{(x=-y)}$ symmetry relates many of the tensor elements in \zref{bts1} so that
 \begin{align}
 \text{Symmetries: }\hat{\sigma}_{z}, \hat{\sigma}_{(x=-y)} \hspace{2cm} \hat T =\mx{
 T_{11} & T_{12} & T_{13} & 0 & 0 & T_{16} \\
 T_{12} & T_{11} & T_{13} & 0 & 0 & T_{16} \\
 T_{31} & T_{31} & T_{33} & 0 & 0 & T_{36} \\
 0 & 0 & 0 & T_{44} & T_{45} & 0 \\
 0 & 0 & 0 & T_{45} & T_{44} & 0 \\
 T_{61} & T_{61} & T_{36} & 0 & 0 & T_{66} }. \label{bts3}
 \end{align}
Finally, having a $\hat{\sigma}_{z}$ and $C_4$ symmetric system we would end up with
 \begin{align}
 \text{Symmetries: }\hat{\sigma}_{z}, C_4\hspace{2cm} \hat T =\mx{
 T_{11} & T_{12} & T_{13} & 0 & 0 & T_{16} \\
 T_{12} & T_{11} & T_{13} & 0 & 0 & -T_{16} \\
 T_{31} & T_{31} & T_{33} & 0 & 0 & 0 \\
 0 & 0 & 0 & T_{44} & T_{45} & 0 \\
 0 & 0 & 0 & -T_{45} & T_{44} & 0 \\
 T_{61} & -T_{61} & 0 & 0 & 0 & T_{66} }. \label{bts4}
 \end{align}
Combining the tensor representations \zref{bts2}-\zref{bts4} one can determine the tensor for an arbitrary combination of the considered point group symmetries. E.g. restoring the tetragonal symmetry would lead to the tensor
 \begin{align}
 \text{Tetragonal symmetry:}\hspace{2cm} \hat T =\mx{
 T_{11} & T_{12} & T_{13} & 0 & 0 & 0\\
 T_{12} & T_{11} & T_{13} & 0 & 0 & 0 \\
 T_{31} & T_{31} & T_{33} & 0 & 0 & 0 \\
 0 & 0 & 0 & T_{44} &0 & 0 \\
 0 & 0 & 0 & 0& T_{44} & 0 \\
 0 & 0 & 0 & 0 & 0 & T_{66} } \label{bts5}.
 \end{align}

\subsection{Setup 1}
In this section we describe the  two proposed experimental setups to detect tetragonal symmetry breaking in elastoresistance measurements in detail. Earlier experiments\cite{Kuo1} glued single crystals of $\text{Ba(Fe}_{1-x}\text{Co}_x\text{)}_2\text{As}_2$ to the surface of a piezo stack as shown in Fig.~\ref{set1}. Varying the strain of the piezo crystal $\epsilon_{xx}, \epsilon_{yy}=-\nu_p \epsilon_{xx}$ (with $\nu_p$ the Poisson ratio of the piezo stack) Kuo et al. measured select admixtures of elastoresistivity coefficients.. Both the resistance and the applied strain are tensors of second rank and can be written in the reducible Voigt notation by  six component arrays
\begin{align}
\begin{split}
\vec R &= (R_{xx}, R_{yy}, R_{zz}, R_{yz},R_{zx}, R_{xy})^T, \\
\bs \epsilon &= (\epsilon_{xx}, \epsilon_{yy}, \epsilon_{zz}, \epsilon_{yz},\epsilon_{zx}, \epsilon_{xy})^T.
\end{split}    \label{s1a}
\end{align}
Defining the relative change of the resistance due to the applied strain as $\delta   r_i = \frac{\delta R_i }{ R_i }$ we can write down the relation
\begin{align}
\delta  r_i = \hat m_{ij} \epsilon_j,    \label{s1b}
\end{align}
with the elastoresistance tensor $\hat m$. Note that although $\hat m$ is a tensor of rank-4 (as it relates two rank-2 tensors), the Voigt notation simplifies this tensor to a matrix form. Consider gluing a rectangular sample, which is cut along the $a,b$ crystal axes, onto the stack with a relative angle of $\theta$ as shown in Fig.~\ref{set1}. The sample is described by the elastoresistance tensor $\hat m$ and the elastic stiffness $\hat C$ (both tensors are in the basis that is shown along the $a,b$ crystal axes). Assuming that we have a system which breaks all of the tetragonal symmetries, we can write down the elastic stiffness and elastoresistance according to \zref{bts1} as\footnote{Note that the elastic stiffness has the more general symmetry $C_{\alpha \beta, \gamma \delta} =C_{\beta\alpha , \gamma \delta}= C_{\alpha \beta,  \delta\gamma}= C_{\gamma \delta,\alpha \beta}$ since it can be written as the second derivative of the free energy with respect to strain: $C_{\alpha \beta, \gamma \delta}= \frac{\partial^2 F}{\partial \epsilon_{\alpha \beta} \partial \epsilon_{\gamma \delta }}$}
 \begin{align}
 \hat C =\mx{
 C_{11} & C_{12} & C_{13} & 0 & 0 & C_{16} \\
 C_{12} & C_{22} & C_{23} & 0 & 0 & C_{26} \\
 C_{13} & C_{23} & C_{33} & 0 & 0 & C_{36} \\
 0 & 0 & 0 & C_{44} & C_{45} & 0 \\
 0 & 0 & 0 & C_{45} & C_{55} & 0 \\
 C_{16} & C_{26} & C_{36} & 0 & 0 & C_{66} } ,  \hspace{1cm} \hat m =  \mx{
 m_{11} & m_{12} & m_{13} & 0 & 0 & m_{16} \\
 m_{21} & m_{22} & m_{23} & 0 & 0 & m_{26} \\
 m_{31} & m_{32} & m_{33} & 0 & 0 & m_{36} \\
 0 & 0 & 0 & m_{44} & m_{45} & 0 \\
 0 & 0 & 0 & m_{54} & m_{55} & 0 \\
 m_{61} & m_{62} & m_{63} & 0 & 0 & m_{66} }  \, . \label{s1c}
 \end{align}
 Upon restoring the tetragonal symmetry, this simplifies to
   \begin{align}
 \hat C =\mx{
 C_{11} & C_{12} & C_{13} & 0 & 0 & 0 \\
 C_{12} & C_{11} & C_{13} & 0 & 0 & 0 \\
 C_{13} & C_{13} & C_{33} & 0 & 0 & 0 \\
 0 & 0 & 0 & C_{44} & 0 & 0 \\
 0 & 0 & 0 & 0 & C_{44} & 0 \\
 0 & 0 & 0 & 0 & 0 & C_{66} },   \hspace{1cm} \hat m =  \mx{
 m_{11} & m_{12} & m_{13} & 0 & 0 &0 \\
 m_{12} & m_{11} & m_{13} & 0 & 0 & 0 \\
 m_{31} & m_{31} & m_{33} & 0 & 0 &0 \\
 0 & 0 & 0 & m_{44} & 0& 0 \\
 0 & 0 & 0 &0& m_{44} & 0 \\
0 & 0 & 0 & 0 & 0 & m_{66} }. \label{s1d}
 \end{align}
Comparing \zref{s1c} with \zref{s1d} we see that, as stated earlier, the breaking of the in-plane mirror symmetries allow for a broad range of new responses (e.g., the elastoresistivity coefficient  $m_{16} =m_{xx,xy} \ne 0$). 
\begin{figure}
 \centering
 \includegraphics[width=0.35\textwidth]{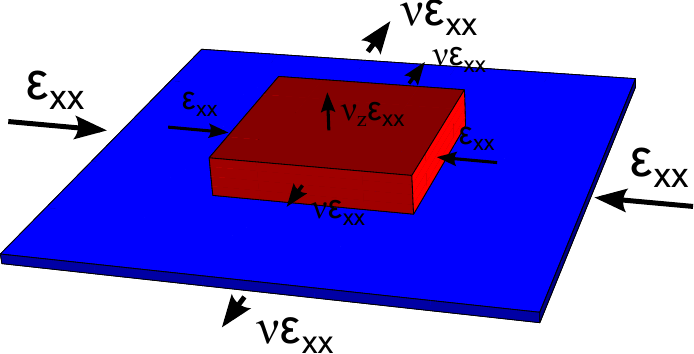} \hspace{1cm}  \includegraphics[width=0.4\textwidth]{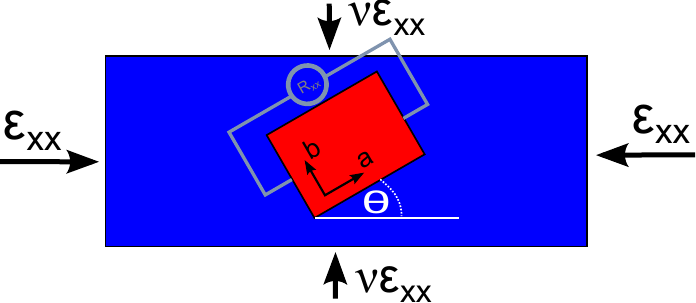}
 \caption{The first example of an experimental setup, where samples which have the same alignment of their crystalline $a-b$ axes are glued at several different angles to a piezoelectric stack.}
 \label{set1}
 \end{figure}

We now consider the relation \zref{s1b} for the rotated sample and in the absence of in-plane mirror symmetries. We  align our coordinate system axes with the rectangular crystal edges and therefore along the  $a,b$ crystal axes. In this system, the elastoresistance and elastic stiffness are given by the expressions in \zref{s1c} and the strain $\bs \epsilon_{\text{crystal}}$ seen by the crystal is related to the strain in the coordinate system of the piezo by a rotation
\begin{align}
\bs \epsilon_{\text{crystal}} &= \hat R  \hat \alpha_\theta \hat R^{-1}  \, \bs \epsilon_{\text{piezo}}  \, .   \label{s1e}
\end{align}
Here, we defined the relation between the real and the engineering strain $\hat R=\text{diag}(1,1,1,2,2,2)$ and the rotation matrix for the 6-component arrays similar to Ref.~[\onlinecite{Kuo1}]:
\begin{align}
\alpha_\theta =  \mx{ \cos ^2(\theta ) & \sin ^2(\theta ) & 0 & 0 & 0 & 2 \cos (\theta ) \sin (\theta ) \\
 \sin ^2(\theta ) & \cos ^2(\theta ) & 0 & 0 & 0 & -2 \cos (\theta ) \sin (\theta ) \\
 0 & 0 & 1 & 0 & 0 & 0 \\
 0 & 0 & 0 & \cos (\theta ) & -\sin (\theta ) & 0 \\
 0 & 0 & 0 & \sin (\theta ) & \cos (\theta ) & 0 \\
 -\cos (\theta ) \sin (\theta ) & \cos (\theta ) \sin (\theta ) & 0 & 0 & 0 & \cos ^2(\theta )-\sin ^2(\theta ) },     \label{s1f}
\end{align}
The strain in the piezo system is given by $\bs \epsilon_{\text{piezo}} =(1,-\nu_p, -\nu_z(\theta),0,0,0) \epsilon_{xx}$. Note, that due to the crystal axis rotation the Poisson ratio $\nu_z(\theta)$ of the crystal in $z$ direction now depends on the rotation angle $\theta$ (see Fig.~\ref{set1}). The angular dependence can be derived by considering the stress-strain relation $\bs \tau_{\text{crystal}} = \hat C \bs \epsilon_{\text{crystal}}$ and the condition $\tau_{3,\text{crystal}} =\tau_{zz,\text{crystal}}=0$ since there is no applied stress in $z$ direction of the crystal. One finds that
\begin{align}
\nu_z(\theta)  = \frac{C_{13} \left(\cos ^2(\theta )-\nu  \sin ^2(\theta )\right)+C_{23} \left(\sin ^2(\theta )-\nu  \cos ^2(\theta )\right)  -C_{36} (\nu +1) \sin (2 \theta )}{C_{33}}.   \label{s1g}
\end{align}
The resistance change $\delta \vec r$ is therefore given by
\begin{align}
\delta  \vec r = \hat m \bs \epsilon_{\text{crystal}} = \hat m \hat R  \hat \alpha_\theta \hat R^{-1}  \, \bs \epsilon_{\text{piezo}},    \label{s1h}
\end{align}
such that the measured $xx$-component is given by
\begin{align}
\delta r_1 (\theta) = \left(\frac{\Delta R}{R}\right)_{xx}\kern-1em(\theta) &=  \biggl( m_{11} \bigl[\cos^2(\theta) -\nu_p \sin^2(\theta)   \bigr]+m_{12} \bigl[\sin^2(\theta) -\nu_p \cos^2(\theta)   \bigr]   \nonumber \\
& \quad -m_{16} (\nu_p+1)  \sin(2\theta) - m_{13} \nu_z(\theta)   \biggr) \cdot \epsilon_{xx}.
\end{align}
Looking at the odd contribution we finally find
\begin{align}
\left(\frac{\Delta R}{R}\right)^{\text{odd}}_{xx}\kern-1em(\theta)&=\frac{1}{2}\left[\left(\frac{\Delta R}{R}\right)_{xx}\kern-1em(\theta)-\left(\frac{\Delta R}{R}\right)_{xx}\kern-1em(-\theta)\right]  =(\nu_p+1)\,\biggl[\frac{m_{xx,zz}C_{zz,xy}}{C_{zz,zz}}-m_{xx,xy}\biggr]\sin(2\theta)\cdot\epsilon_{xx}.\nonumber
\end{align}

\subsection{Setup 2}
The second proposed setup is similar to the one described in the previous section, but here we start from a thin film crystal and use lithography to cut out the rectangular sections as shown in Fig.~\ref{set2}. The advantage of this method is that aligning the crystal by gluing it on the piezo element with high precision is a rather difficult task, whereas in using lithography, one can cut out the samples very precisely.  
\begin{figure} [h]
 \centering
 \includegraphics[width=0.8\textwidth]{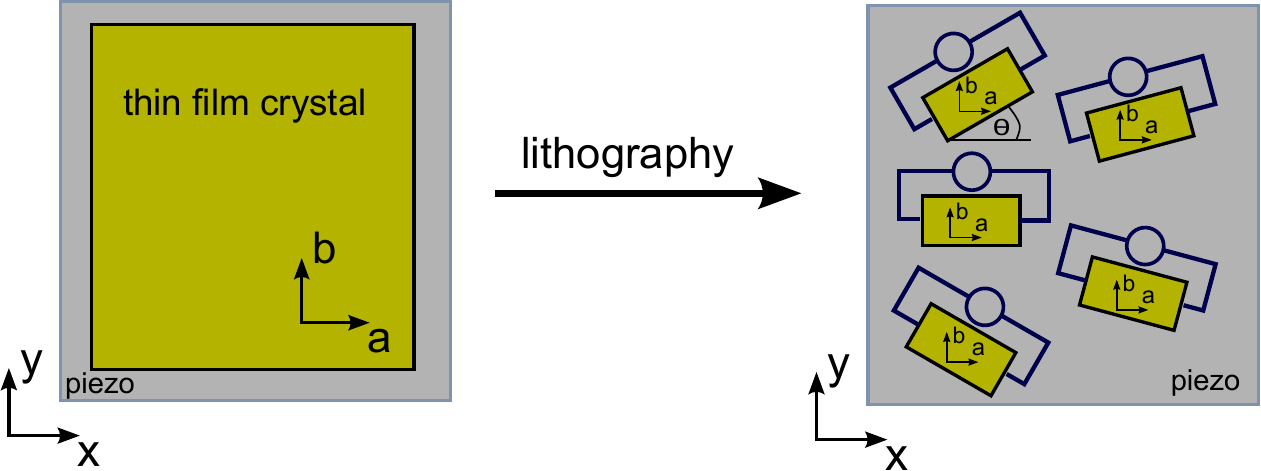}
 \caption{A second possible experimental setup, where crystals with different alignment of their crystalline axes can  onto the piezoelectric stack.}
 \label{set2}
 \end{figure}
In contrast to the other experiment, the crystal axes $a,b$ are aligned with the principal axes of the piezo stack. Therefore, the Poisson ratio in $z$ direction does not depend on the angle of the cut. If we consider the coordinate system to be along the $a,b$ (and thus also $x,y$) axis, what is measured is effectively the rotated resistance
\begin{align}
\delta \vec r(\theta) = \hat \alpha_\theta \delta \vec r_0 =  \hat \alpha_\theta \hat m \cdot \bs \epsilon_{\text{piezo}} .
\end{align}
The measured longitudinal resistance is therefore given by
\begin{align}
\delta r_1 (\theta) = \ \left(\frac{\Delta R}{R}\right)_{xx}\kern-1em(\theta)&=  \biggl( \cos^2(\theta) \bigl[m_{11} - \nu m_{12} - \nu_z m_{13}   \bigr] +\sin^2(\theta) \bigl[m_{21} - \nu m_{22} - \nu_z m_{23}   \bigr]     \nonumber \\
& \qquad    -\sin(2 \theta) \bigl[m_{61} - \nu m_{62} - \nu_z m_{63}   \bigr] \biggr) \epsilon_{xx} ,
\end{align}
and the antisymmetric part 
\begin{align}
 \left(\frac{\Delta R}{R}\right)^{\text{odd}}_{xx}\kern-1em(\theta) =   -\sin(2 \theta) \bigl[m_{61} - \nu m_{62} - \nu_z m_{63}   \bigr] \epsilon_{xx} ,
\end{align}
is again proportional to $\sin(2\theta)$ and only present for broken mirror symmetries.

\end{widetext}

%\begin{thebibliography}{99}

%\end{thebibliography}

\end{document}